\documentclass[12pt, draftclsnofoot, onecolumn]{IEEEtran}
\usepackage{cite}

\usepackage{graphicx}

\ifCLASSINFOpdf
\else
\fi
\usepackage{color}
\usepackage{graphicx, subfigure}

\usepackage[cmex10]{amsmath}

\usepackage{amsfonts}

\begin{document}
%
\title{Generalized Multicast Multibeam Precoding for Satellite Communications}
%
%
%

\author{Vahid Joroughi,
        Miguel \'Angel V\'azquez
				and~Ana I. P\'erez-Neira,~\IEEEmembership{Senior Member,~IEEE}
\thanks{The research leading to these results has received funding from the Spanish Ministry of Science and Innovation under projects TEC2014-59225-C3-1-R (ELISA) and the Catalan Government (2014 SGR 1567).}
\thanks{V. Joroughi and A. P\'erez-Neira are with the Universitat Politecnica de Catalunya (UPC) and Centre Tecnol\`ogic de les Telecomunicacions de 
Catalunya (CTTC), Barcelona, Spain.}
\thanks{Emails:vahid.joroughi@upc.edu,~ana.isabel.perez@upc.edu}
\thanks{M. \'A. V\'azquez is with the Centre Tecnol\`ogic de les Telecomunicacions de 
Catalunya (CTTC), Barcelona, Spain.}
\thanks{Emails:mavazquez@cttc.es}}

\maketitle

\begin{abstract}
This paper deals with the problem of precoding in multibeam satellite systems. In contrast to general multiuser multiple-input-multiple-output (MIMO) cellular schemes, multibeam satellite architectures suffer from different challenges. First, satellite communications standards embed more than one user in each frame in order to increase the channel coding gain. This leads to the different so-called multigroup multicast model, whose optimization requires computationally complex operations. Second, when the data traffic is generated by several Earth stations (gateways), the precoding matrix must be distributively computed and attain additional payload restrictions. Third, since the feedback channel is adverse (large delay and quantization errors), the precoding must be able to deal with such uncertainties. In order to solve the aforementioned problems, we propose a two-stage precoding design in order to both limit the multibeam interference and to enhance the intra-beam minimum user signal power (i.e. the one that dictates the rate allocation per beam). A robust version of the proposed precoder based on a first perturbation model is presented. This mechanism behaves well when the channel state information is corrupted. Furthermore, we propose a per beam user grouping mechanism together with its robust version in order to increase the precoding gain. Finally, a method for dealing with the multiple gateway architecture is presented, which offers high throughputs with a low inter-gateway communication. The conceived designs are evaluated in a close-to-real beam pattern and the latest broadband communication standard for satellite communications.
\end{abstract}

\begin{IEEEkeywords}
Multibeam satellite systems, Precoding, Robust design, Multigroup multicast.
\end{IEEEkeywords}

%
\IEEEpeerreviewmaketitle

\section{Introduction}

\subsection{Motivation}

\IEEEPARstart{S}{atellite} communications will play a central role towards fulfilling next generation 5G communication requirements \cite{5G}. As a matter of fact, anytime-anywhere connectivity cannot be conceived without the presence of the satellite segment \cite{5Gs}. Indeed, the satellite  industry is not only targeting areas without backbone connectivity (maritime, aeronautic, rural), but also high dense populated scenarios with an existing communication infrastructure, where the satellite will become an essential element to decongest the terrestrial wireless network.

In order to deliver broadband interactive data traffic, satellite payloads are currently implementing a multibeam radiation pattern. The use of a multibeam architecture brings several advantages in front of a single global beam transmission \cite{SAT:SAT1049}. First, since an array fed reflector is employed, the antenna gain to noise ratio can be increased leading to high gain of each beam return link achievable throughput. Second, different symbols can be simultaneously sent to geographically separated areas, allowing a spatially multiplexed communication. Last but not least, the available bandwidth can be reused in sufficiently separated beams, leading to an increase of the user bandwidth yet maintaining a low multiuser interference. 

Nevertheless, whenever the system designers target the terabit satellite system (i.e. a satellite system offering a terabit per second capacity), the aforementioned multibeam architecture shall be reconsidered. Precisely, full frequency reuse among beams becomes mandatory in order to support the terabit capacity as larger available user bandwidth is required. As a consequence, when considering the satellite forward link, interference mitigation techniques need to be implemented either at the user terminal (multiuser detection) or in the transmitter (precoding).

Whenever precoding is employed, the users must feed back their channel station information (CSI) to the transmitter so that it can revert the interference effect at the transmit side. These techniques relay severely on the quality of the CSI and they dramatically decrease their performance in case CSI is either deprecated or corrupted. On the contrary, multiuser detection techniques does not depend on the feedback channel but; however, the user terminal complexity increases so as its cost. In addition, precoding system level studies are providing encouraging results for implementing this technique in real multibeam satellite systems \cite{daniel}. As a result, we will consider precoding as interference reliever for the present study.

\subsection{Previous Works}

The first designs of precoding techniques for multibeam satellite systems can be found in \cite{EURECOM06}. Mimicking the linear techniques for multiuser multiple-input-multiple-output (MIMO) schemes the authors propose a zero forcing (ZF) and minimum mean square error (MMSE) precoding designs. In addition, several challenges for the implementation of precoding in multibeam satellite systems were pointed out. We describe them in the following.

As a general statement, the payload complexity shall be maintained low and; consequently, the ground segment must perform most of the computations and transmit the precoded signals through the feeder link. This feeder link must be able to support the overall satellite traffic, leading to a very large bandwidth requirement. This requirement is even larger when precoding  is deployed at the gateway since the feed signals must be precoded and simultaneously transmitted through the feeder link. On the other hand, if the payload is equipped with a beamforming mechanism, the feeder link bandwidth requirements can be alleviated since only the user signals shall be transmitted\footnote{This statement assumes that the number of feeds is larger than the number of users as it happens in most of the current deployments. This will be discussed in Section II.}. However, as presented in \cite{suso12,6133895} on board processing limits the overall gains obtained by the on ground processing \cite{zheng12,unilu12}. Remarkably, in case the on board processing is optimized considering an underlying precoding scheme, certain throughput gain can be preserved \cite{vahid13}. Yet another alternative is to shift the feeder link to the Q/V bands although certain diversity schemes must be deployed for dealing with the large path loss \cite{6995937}.

In order to deal with the extremely increase of feeder link bandwidth requirements resulting from the full frequency beam reuse pattern, several Earth stations (gateways) can be deployed. With this, the available spectrum for the feeder link can be reused among spatially separated gateways through very directive antennas. In these systems, the traffic is generated at different gateways so that several feeder links simultaneously transmit precoded data from isolated areas. As a consequence, the precoding technique shall be reconsidered since, in order to have enough spectrum to access all the feeds, each gateway has only access to a certain set of feeds. In \cite{6190334,6843130} a precoding scheme based on the regularized ZF scheme is presented. As a matter fact, the overall throughput is reduced when multiple gateways are considered even though computationally complex on ground schemes are deployed \cite{6362558}.

Last, but not least, satellite communications operate in a multicast fashion since data from different users is embedded in the same frame. Precisely, in order to increase the coding gain, each beam simultaneously serves more than one user by means of transmitting a single coded frame. This scheme entails a modification of the overall precoding scheme since more than one spatial signature per beam must be considered and; moreover, the achievable rate is dictated by the user with the lowest signal-to-noise-plus-interference ratio (SNIR). The first approach for solving this problem can be found in \cite{4686703} were a sequential beamforming scheme is presented. However, whenever a large number of beams over the coverage area is targeted, one-shot scheme must be implemented. An example of this can be found in \cite{6843054} and \cite{6934558} where a design based on MMSE precoding and channel averaging is presented. Since the multicast multibeam transmission can be cast as multigroup multicast mathematical model, the proposed designs in \cite{4840357} can be applied. Finally, very high throughputs can be obtained whenever the joint precoding and user scheduling is performed as in \cite{7091022}.

\subsection{Contributions}

In contrast to the aforementioned works, the present paper proposes a low complex ground precoding scheme to deal with the multibeam interference in multicast transmissions. The presented novel technique is based on two stage linear precoding similar to \cite{Stankovic08}, where the multiuser MIMO scenario is targeted. Our proposal offers higher spectral efficiencies than the regularized ZF \cite{4840357} and the average MMSE scheme \cite{6843054,6934558} yet maintaining a low computational complexity. It is important to remark that we prioritize low complex one-shot design in front of iterative gradient-based schemes \cite{4840357,6362558}. 

In addition, considering that the CSI will be corrupted at the gateway, a robust scheme is presented based on the first order perturbation theory of the eigenvectors and eigenvalues \cite{Liu08}. This robust design is novel and it has not been applied before. The resulting precoding design remains low complex so that it can be implemented even if a very large number of feeds are considered.

Since the achievable rates decrease whenever the user channel vectors within one beam are not collinear, we propose a user grouping technique. With this, over all possible users to be served for each beam, we select the most adequate ones using a variation of the $k$-means algorithm. This algorithm differs from the one presented in \cite{6843054} as not only the channel magnitudes but the phase effects are considered. Additionally, a novel robust user grouping scheme is proposed in order to deal with the possible channel uncertainties.

In case the data traffic is generated by several gateways, a precoding mechanism is presented for dealing with the main challenges; namely, CSI sharing and the distributed precoding matrix computation. Both a reduced inter-gateway communication for CSI sharing and a precoding matrix division among gateways are presented. Even though the achievable rates are decreased when the multiple gateway architecture is considered, the proposed scheme offers a good trade-off between communication overhead, payload complexity and overall throughput.

The rest of the paper is organized as follows. Section II introduces the problem so as the channel modelling. Based on this model and problem statement, Section III presents the two stage precoding design for dealing with both the multibeam interference and intra-beam signal enhancement.  Relaying on a certain precoding design, a robust scheme is proposed based on the first order perturbation method in Section IV. Section V shows how to implement precoding in a multiple gateway architecture. Section VI presents numerical simulations considering the digital video broadband S2X (DVB-S2X) and a beam pattern of 245 beams. Section VII concludes.

%
%
%
%


\textit{Notation:} We adopt the notation of using lower case boldface for vectors, $\mathbf{v}$, and upper case boldface for matrices, $\mathbf{A}$. The transpose operator and the conjugate transpose operator are denoted by the symbols $(\cdot)^T$, $(\cdot)^H $ respectively. $\mathbf{E}[\cdot]$ denotes expectation. $\mathbf{I}$ denotes the identity matrix. $\mathbb{C}$ denotes the complex numbers. $\| \cdot \|$ denotes the Euclidean norm. $| \cdot |$ denotes the absolute value. $\preceq$ denotes the componentwise inequality. $\circ$ denotes the Hadamard product.

\section{Problem Formulation}

\subsection{Channel Model}

Let us consider a multibeam broadband satellite with fixed receivers where a single gateway is provisioning signals to be transmitted through a feeder link. Over the feeder link\footnote{The feeder link connection is considered ideal. This is, noiseless and without channel impairments.}, a total number of $N$ feed signals are frequency multiplexed so that the payload has to detect and route them through an array fed reflector. This array fed reflector transforms the $N$ feed signals into $K$ transmitted signals (i.e. one signal per beam) to be radiated over the multibeam coverage area. 

As a matter of fact, the array fed reflector can have a single-feed-per-beam (SFPB) architecture whenever $N = K$ or a multiple-feed-per-beam (MFPB) when $N > K$. This latter payload architecture presents lower beamforming scan losses and larger antenna gains than the SFPB \cite{mfpb2011}. For the sake of generality will consider the MFPB structure in the rest of the paper.

\begin{figure}[h!]
\vspace{-1.3 cm}
  \centering
    \includegraphics[width=14.5cm, height=10cm]{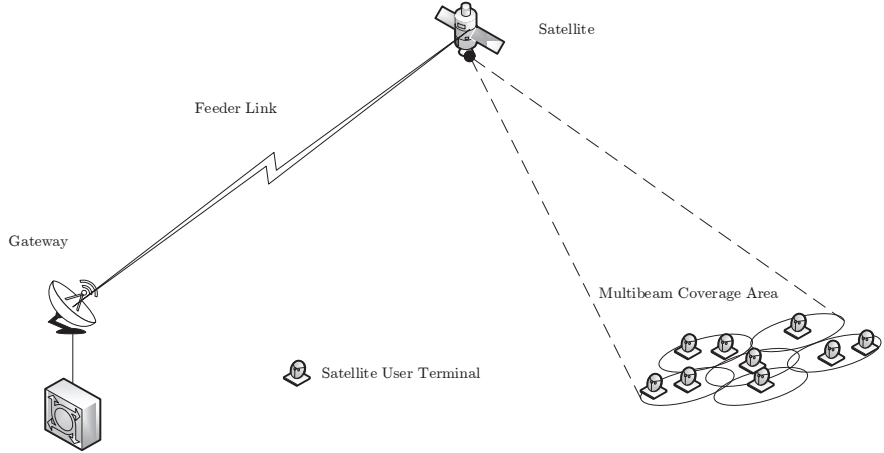}
		\vspace{-2.6 cm}
      \caption{The picture depicts the multicast multibeam satellite structure. The gateway delivers certain data to the coverage area by first the feeder link and; posteriorly, the satellite. While the feeder link multiplexes $N$ signals, the satellite that is equipped with an array fed reflector, radiated a total of $K$ signals (one signal per beam). Every radiated signal by the satellite shall be detected by a total number of $Q$ users per beam.}
      \label{system}
\end{figure}

Towards a spectrally efficient communication, all beams share the frequency band and; in a given time instant, the $k$-th beam simultaneously serves a total amount of $Q_k$ users. In other words, in a give time instant the scheduler selects a set of $Q_k$ users at the $k$-th beam (same for all beams) and it constructs a codeword with information to be transmitted to all $Q_k$. Without loss of generally, we will assume that each beam serves the same number of users simultaneously and it is equal to $Q$ (i.e. $Q_k=Q \quad  k=1,\ldots,K$). Figure \ref{system} shows the overall system.

Under this context, the received signal can be modelled as
\begin{equation}
\mathbf{y} = \mathbf{H}\mathbf{x} + \mathbf{n}
\end{equation}
where $\mathbf{y} \in \mathbb{C}^{KQ \times 1}$ is a vector containing the received signals at each user terminal. Vector $\mathbf{n} \in \mathbb{C}^{KQ \times 1}$ contains the noise terms of each user terminal and we will assume that they are Gaussian distributed with zero mean, unit variance and uncorrelated with both the desired signal and the other users noise terms (i.e. $E\left[\mathbf{n}\mathbf{n}^H\right] = \mathbf{I}_{KQ}$). 

The channel matrix can be described as follows
\begin{equation}
\mathbf{H} = \mathbf{F}  \circ \mathbf{\Phi}.
\label{kakh}
\end{equation}
Matrix $\mathbf{F} \in \mathbb{R}^{KQ \times N}$ represents signal attenuation generated via both the atmospheric fading and the antenna feed radiation. This matrix can be decomposed as follows
\begin{equation}
\mathbf{F} = \mathbf{A}\mathbf{G}
\end{equation}
where $\mathbf{A} \in  \mathbb{R}^{KQ \times KQ}$ is diagonal  matrix whose diagonal entries are the atmospheric fading terms corresponding to the $q$-th user in the $k$-th beam. Matrix $\mathbf{G}\in  \mathbb{R}^{KQ \times N}$ takes into account the rest of gain and loss factors. Its $(kq,n)$-th entry can be described as follows
\begin{equation}
\left[G \right]_{k,n} = \frac{G_Ra_{kqn}}{4 \pi \frac{d_{kq}}{\lambda} \sqrt{K_BT_RB_W}}
\end{equation}
with $d_{kqn}$ the distance between the $q$-th user terminal in the $k$-th beam and the satellite. $\lambda$ is the carrier wavelength, $K_B$ is the Boltzmann constant, $B_W$ is the carrier bandwidth, $G_R^2$ the user terminal receive antenna gain, and $T_R$ the receiver noise temperature. The term $a_{kqn}$ refers to the gain from the $n$-th feed to the $q$-th user in the $k$-th beam. It is important to mention that the $\mathbf{G}$ matrix has been normalized to the receiver noise term.

Furthermore, matrix $\mathbf{\Phi} \in \mathbb{C}^{KQ \times N}$ represents the phase variation effects between the $n$-th feed signal and the $q$-th user located in the $k$-th beam. Each entry is defined as
\begin{equation}
\left[\mathbf{\Phi} \right]_{q,n} = e^{j\theta_{q,n}} 
\end{equation}
where $\theta_{q,n}$ is uniformly distributed between 0 and 2$\pi$.

The aforementioned phase effect determines the performance of the system severely. Due to that, an option is to use ultra stable oscillators so that the phase variation due to the different feed oscillators is reduced. With this, we can assume another phase distribution such that
\begin{equation}
\theta_{q,n}^{ultra} = \phi_q + \gamma_{q,n}
\end{equation}
where $\phi_q$ is a uniformly distributed in 0 and 2$\pi$ and $\gamma_{q,n}$ is modelled as a zero mean Gaussian distribution with variance $\chi$.

For notation convenience, it is better to define $\mathbf{H}_k \in \mathbb{C}^{Q \times N}$ as the channel matrix for the $k$-th beam so that
\begin{equation}
\mathbf{H} = \left[\mathbf{H}_1^T, \mathbf{H}_2^T, \ldots , \mathbf{H}_K^T \right]^T.
\end{equation}
Moreover, the channel vector of the $q$-th user in the $k$-th beam is defined as $\mathbf{h}_{k,q}$ so that
\begin{equation}
\mathbf{H}_k = \left[\mathbf{h}_{k,1}^T, \mathbf{h}_{k,2}^T, \ldots , \mathbf{h}_{k,Q}^T \right]^T.
\end{equation}

In order to minimize the multiuser interference resulting from the full frequency reuse, linear precoding is considered. Under this context, the transmitted symbol can be modelled as
\begin{equation}
\mathbf{x} = \mathbf{W}\mathbf{s}
\end{equation}
where $\mathbf{s} \in \mathbb{C}^{K \times 1}$ is a vector that contains the transmitted symbols which we assume uncorrelated and unit norm $\left(E\left[\mathbf{s}\mathbf{s}^H \right] = \mathbf{I}_{K} \right)$. Matrix $\mathbf{W} \in \mathbb{C}^{N \times K}$ is the linear precoding matrix to be designed.

\subsection{Precoding Design}

Let us formulate the precoding design of a multicast multibeam satellite system. The overall system performance can be optimized considering the maximum sum rate:
\begin{equation}\label{sumrate}
\begin{aligned}
& \underset{\mathbf{W}}{\text{maximize}} \quad  \sum_{k=1}^K\underset{q = 1, \ldots Q}{\text{minimum}} \quad  r_{k,q} \\
& \text{subject to}\\
&\text{Tr}\left(\mathbf{W}\mathbf{W}^H\right) \leq {P_T} \\
\end{aligned}
\end{equation}
where $r_{q,k}$ denotes the achievable rate of the $q$-th user at the $k$-th beam,
\begin{equation}
r_{k,q} = \log_2\left(1 + \frac{|\mathbf{h}_{q,k}^H\mathbf{w}_k|^2}{\sum_{j \neq k}^K|\mathbf{h}_{q,k}^H\mathbf{w}_j|^2 + \sigma^2 } \right),
\end{equation}
$P_T$ denotes the available power at the satellite and $\mathbf{w}_k$ corresponds to the $k$-th column of matrix $\mathbf{W}$.  It is important to remark that even though power sharing mechanisms among beams can be implemented \cite{multi07, flepowe} their deployment into next generation satellite payloads will require costly and complex radio-frequency designs. Under this context, the precoding design shall consider a per-feed available power constraint so that the constraint in \eqref{sumrate} becomes
\begin{equation}
\left(\mathbf{W}\mathbf{W}^H\right)_{n,n} \leq \frac{P_T}{N} \quad n=1, \ldots , N,
\end{equation}
where it has been assumed that the available power is equally share by all feed elements.
 
In any case, problem \eqref{sumrate} is a difficult non-convex problem whose convex relaxations even require computationally demanding operations \cite{1574196}. For multibeam satellite systems the computational complexity of the precoding design is an essential feature since these systems usually operate with hundreds of beams. Consequently, in contrast to other works, the target of this paper is to design a low computationally complex precoding scheme able to achieve high throughput values. This is presented in the next section.

\section{Generalized Multicast Multibeam Precoding}

The precoding design in multicast multibeam satellite systems has to main roles. First, the inter-beam interference shall be minimized and; second, the precoding shall increase the lowest SINR within each beam. Under this context, the precoding design can be divided into two sub-matrices such as
\begin{equation}
\mathbf{W} = \alpha\mathbf{W}_a\mathbf{W}_b,
\end{equation}
where
\begin{equation}
\mathbf{W}_a = \left[\mathbf{W}_{a_1}, \ldots,  \mathbf{W}_{a_K}\right] \in \mathbb{C}^{N \times KQ},
\end{equation}
and
\begin{equation}
\mathbf{W}_b = 
\begin{bmatrix}
  \mathbf{w}_{b_1} & \mathbf{0} & \cdots & \mathbf{0} \\
  \mathbf{0} & \mathbf{w}_{b_2} & \cdots & \mathbf{0} \\
  \vdots  & \vdots  & \ddots & \vdots  \\
  \mathbf{0} & \mathbf{0} & \cdots & \mathbf{w}_{b_K}
 \end{bmatrix}
 \in \mathbb{C}^{KQ \times K},
\end{equation}
with $\mathbf{W}_{a_k} \in \mathbb{C}^{N \times Q}$ and $\mathbf{w}_{b_k} \in \mathbb{C}^{Q \times 1}$. The matrix $\mathbf{W}_a$ is used to mitigate the inter-beam interference first and then the matrix $\mathbf{W}_b$ is used to optimize the intra-frame data rate (i.e. the rate of the served users) so that $\mathbf{w}_{b_k}$ and $\mathbf{W}_{a_k}$ denotes the precoding for optimizing the rate at $k$-th beam. Finally, the parameter $\alpha$ is chosen to set the available power constraint (both for the per feed and total power constraint).

In the following subsections, two different designs for  $\mathbf{W}_a$ and a single design for $\mathbf{W}_b$ are presented.

\subsection{Inter-beam Interference Mitigation Precoding}

\subsubsection{Multibeam Interference Mitigation (MBIM)}

Let us define $\widetilde{\mathbf{H}}_k$ as
\begin{equation}
\widetilde{\mathbf{H}}_k = 
\begin{bmatrix}
\mathbf{H}_1 \\
\vdots \\
\mathbf{H}_{k-1} \\
\mathbf{H}_{k+1} \\
\vdots \\
\mathbf{H}_K
\end{bmatrix}
\in \mathbb{C}^{(K-1)Q \times N}.
\end{equation}

First we observe the interference impact of the precoding matrix $\mathbf{W}_a$. This can be done by means of constructing the equivalent combined channel matrix after the precoding effect:
\begin{equation}
\mathbf{H}\mathbf{W}_a = 
\begin{bmatrix}
  \mathbf{H}_1\mathbf{W}_{a_1} & \mathbf{H}_1\mathbf{W}_{a_2} & \cdots & \mathbf{H}_1\mathbf{W}_{a_K} \\
  \mathbf{H}_2\mathbf{W}_{a_1} & \mathbf{H}_2\mathbf{W}_{a_2} & \cdots & \mathbf{H}_2\mathbf{W}_{a_K} \\
  \vdots  & \vdots  & \ddots & \vdots  \\
  \mathbf{H}_K\mathbf{W}_{a_1} & \mathbf{H}_K\mathbf{W}_{a_2} & \cdots & \mathbf{H}_K\mathbf{W}_{a_K}
 \end{bmatrix},
\end{equation}
where the $k$-th beam effective channel is given by $\mathbf{H}_k\mathbf{W}_{a_k}$ and the interference generated to the other users is determined by $\widetilde{\mathbf{H}}_k\mathbf{W}_{a_k}$.

As described in \cite{Stankovic08} an efficient design of $\mathbf{W}_{a}$ is given by the optimal matrix of the following modified MMSE objective function
\begin{equation}\label{mmse}
\begin{aligned}
\underset{\mathbf{W}_a}{\text{minimize}} \quad  \text{E} \left[\sum_{k=1}^K \|\widetilde{\mathbf{H}}_k\mathbf{W}_{a_k}  \|^2 + \frac{KQ}{P_{T}}\|\mathbf{n} \|^2 \right].
\end{aligned}
\end{equation}
where the term $\frac{KQ}{P_{T}}$ is obtained considering a total power constraint.

The solution of this optimization problem is given by
\begin{equation}
\mathbf{W}_{a_k}^{\text{MBIM}} = \mathbf{M}_{a_k}\mathbf{D}_{a_k},
\end{equation}
where
\begin{equation}
\mathbf{M}_{a_k} = \widetilde{\mathbf{V}}_k,
\end{equation}
and
\begin{equation}
\mathbf{D}_{a_k} = \left(\widetilde{\boldsymbol{\Sigma}}_k + \frac{KQ}{P_{T}}\mathbf{I} \right)^{-1/2}.
\end{equation}
Note that it has been considered the singular value decomposition of $\widetilde{\mathbf{H}}_k^H\widetilde{\mathbf{H}}_k = \widetilde{\mathbf{V}}_k\widetilde{\boldsymbol{\Sigma}}_k\widetilde{\mathbf{V}}_k^H$.

\subsubsection{Regularized Zero-Forcing}

Let $\mathbf{H}^{(R)} \in \mathbb{C}^{KQ \times KQ}$ denote the regularized channel defined as
\begin{equation}
\mathbf{H}^{(R)} = \mathbf{H}\mathbf{H}^H + \frac{KQ}{P_{T}}\mathbf{I},
\end{equation}
where the same regularization factor as that of the multicast-aware regularized zero-forcing \cite{Sil09} is considered.
Let us define $\widetilde{\mathbf{H}}_k^{(R)}$ as
\begin{equation}
\widetilde{\mathbf{H}}_k^{(R)} = 
\begin{bmatrix}
\widetilde{\mathbf{H}}_1^{(R)} \\
\vdots \\
\widetilde{\mathbf{H}}_{k-1}^{(R)} \\
\widetilde{\mathbf{H}}_{k+1}^{(R)} \\
\vdots \\
\widetilde{\mathbf{H}}_K^{(R)}
\end{bmatrix}
\in \mathbb{C}^{K(Q-1) \times KQ}.
\end{equation}
The SVD decomposition of matrix can be described as
\begin{equation}
\widetilde{\mathbf{H}}_k^{(R)} = \widetilde{\mathbf{U}}_k^{(R)}\widetilde{\boldsymbol{\Sigma}}_k^{(R)} \left[\widetilde{\mathbf{V}}_k^{(R),1},\widetilde{\mathbf{V}}_k^{(R),0} \right]^H,
\end{equation}
where it is emphasized that there is always a null space of dimension $Q$ spanned by $\widetilde{\mathbf{V}}_k^{(R),0}$. This matrix is used for this scheme such as
\begin{equation}
\mathbf{W}_{a_k}^{\text{R-ZF}} = \mathbf{H}^H\widetilde{\mathbf{V}}_k^{(R),0} \in \mathbb{C}^{N \times Q}.
\end{equation}

\subsection{Intra-beam Precoding}

After the first precoding scheme, $\mathbf{W}_a$, the $k$-th beam observes an equivalent channel $\mathbf{Z}_k = \mathbf{H}_k\mathbf{W}_{a_k} \in \mathbb{C}^{Q \times N}$. Based on $\mathbf{Z}_k$ the system designer shall construct $\mathbf{w}_{b_k}$. Let us mention that the equivalent channel for the $q$-th user located at the $k$-th beam is denoted by $\mathbf{z}_{k_q} \in \mathbb{C}^{N \times 1}$ so that
\begin{equation}
\mathbf{Z}_k = \left[\mathbf{z}_{k_1}, \ldots, \mathbf{z}_{k_Q} \right].
\end{equation}

A suboptimal yet efficient approach is to maximize the average SNR considering the equivalent channel matrix. This is done with the following optimization problem
\begin{equation}\label{maxmeanwb}
\begin{aligned}
& \underset{\mathbf{w}_{b_k}}{\text{maximize}} \quad  \sum_{q=1}^Q |\mathbf{z}_{k_q}^H\mathbf{w}_{b_k}|^2 \\
& \text{subject to}\\
&\|\mathbf{w}_{b_k}\|^2 \leq 1.  \\
\end{aligned}
\end{equation}
Since the objective function in \eqref{maxmeanwb} can be re-written as $\|\mathbf{Z}_{k}\mathbf{w}_{b_k}\|^2$, the optimal solution of \eqref{maxmeanwb} is given by the eigenvector associated to the largest eigenvalue of matrix $\mathbf{Z}_{k}$. Note that this precoding design offers a low computational complexity since only the eigenvector associated to the largest eigenvalue needs to be computed. 

\subsection{Power control}

Once both precoding schemes are computed, it is time to calculate the value of $\alpha$ in order to fulfil the maximum transmit power constraints. In case the the maximum per feed available power constraint is considered, 
\begin{equation}
\alpha = \frac{P_T}{N \text{max}_{i}\left(\left(\mathbf{W}\mathbf{W}^H \right)_{i,i}\right)},
\end{equation}
whereas in case total power constraints are considered
\begin{equation}
\alpha = \frac{P_T}{N \text{Tr}\left(\left(\mathbf{W}\mathbf{W}^H \right)_{i,i}\right)}.
\end{equation}

\section{Robust Multicast Multibeam Precoding}

As it will be shown in the simulation section, the MBIM schemes offers larger achievable rates than the R-ZF. Consequently, we will consider this design so as the average optimization intra-beam method. In any case, note that precoding performance relies on an accurate CSI fed back by the receiver. However, this CSI suffers from certain degradation due to quantization and transmission delay. Due to that, it is convenient to reformulate the optimization problem in order to take into account these possible variations \cite{1561600}. This is done in the following subsections for both the inter-beam and intra-beam precoding.

First, let us introduce the perturbation where the transmitter do not longer have access to $\mathbf{H}$ but to a degraded version such as
\begin{equation}
\widehat{\mathbf{H}} = \mathbf{H} + \boldsymbol{\Delta},
\label{kakh2}
\end{equation}
where $\boldsymbol{\Delta} \in \mathbb{C}^{K \times N}$ is the perturbation matrix where it is assumed to be constrained so that
\begin{equation}
\|\boldsymbol{\Delta}\|^2 \leq \gamma.
\end{equation}
For notational convenience, it is important to define the following sub-matrices
\begin{equation}
\boldsymbol{\Delta} = \left[\boldsymbol{\Delta}_1^T, \boldsymbol{\Delta}_2^T, \ldots , \boldsymbol{\Delta}_K^T \right]^T,
\end{equation}
\begin{equation}
\boldsymbol{\Delta}_k = \left[\boldsymbol{\delta}_{k,1}^T, \boldsymbol{\delta}_{k,2}^T, \ldots , \boldsymbol{\delta}_{k,Q}^T \right]^T,
\end{equation}
\begin{equation}
\widetilde{\boldsymbol{\Delta}}_k = 
\begin{bmatrix}
\boldsymbol{\Delta}_1 \\
\vdots \\
\boldsymbol{\Delta}_{k-1} \\
\boldsymbol{\Delta}_{k+1} \\
\vdots \\
\boldsymbol{\Delta}_K
\end{bmatrix}
\in \mathbb{C}^{(K-1)Q \times N},
\end{equation}
where $\boldsymbol{\Delta}_k \in \mathbb{C}^{Q \times N}$ is the perturbation associated to the $k$-th beam, $\boldsymbol{\delta}_{k,q}\in \mathbb{C}^{N \times 1}$ is the perturbation associated to the $q$-th user located at the $k$-th beam. Under this context, the perturbation sub-matrices can be constrained as follows
\begin{equation}
\|\boldsymbol{\Delta}_k\|^2 \leq \gamma_k,
\end{equation}
\begin{equation}
\|\widetilde{\boldsymbol{\Delta}}_k\|^2 \leq  \widetilde{\gamma_k} ,
\end{equation}
for $k = 1 \ldots , K$ and where
\begin{equation}
\widetilde{\gamma_k} =  \sum_{l \neq k}^K \gamma_l.
\end{equation}

Additionally, a definition of a  lower bound of the perturbation matrices $\boldsymbol{\Delta}_k$ is convenient
\begin{equation}
\|\boldsymbol{\Delta}_k\|^2 \geq \underline{\gamma_k}.
\end{equation}

Remarkably, finding the adequate $\gamma$ bounds for all different perturbation matrices is not an easy task. Indeed, the computation of the different bounds shall be done on a empirical basis considering the different error sources and their final value on the perturbation matrix. This study is out of the scope of this paper and we will only provide a sensitivity analysis in the simulation section.

\subsection{Robust Inter-beam Precoding}

Considering the MBIM scheme presented in the previous section, whenever the robust worst-case optimization problem \eqref{mmse} is targeted, the following new optimization problem shall be considered
\begin{equation}\label{rmmse}
\begin{aligned}
& \underset{\mathbf{W}_a}{\text{minimize}} \quad \underset{\{\widetilde{\boldsymbol{\Delta}}_k \}_{k=1}^K}{\text{maximize}} \quad  \text{E} \left[\sum_{k=1}^K \|\widehat{\widetilde{\mathbf{H}}}_k\mathbf{W}_{a_k}  \|^2 + \frac{KQ}{P_{T}}\|\mathbf{n} \|^2 \right]\\
& \text{subject to}\\
& \|\widetilde{\boldsymbol{\Delta}}_k \|^2 \leq \widetilde{\gamma_k} \quad k=1, \ldots , K, \\
\end{aligned}
\end{equation}
where $\widehat{\widetilde{\mathbf{H}}}_k = \widetilde{\mathbf{H}}_k + \widetilde{\boldsymbol{\Delta}}_k$. Considering that the optimal design on \eqref{mmse} leads to the computation of an eigendecomposition, the perturbation matrix will both impact the eigenvectors and eigenvalues. Robust designs generally only consider the effect on the eigenvalues \cite[Chapter 7]{palomar}; however, the impact on the eigenvectors cannot be considered negligible \cite{Liu08}.

The following theorem provides an approximate solution of the optimization problem in \eqref{rmmse}. 

\textbf{Theorem 1} \emph{The optimal inter-beam precoding matrix which approximately minimizes \eqref{rmmse} is
\begin{equation}
\widehat{\mathbf{W}}_{a_k}^{\text{MBIM}} = \widehat{\mathbf{M}_{a_k}}\widehat{\mathbf{D}_{a_k}},
\end{equation}
where
\begin{equation}
\widehat{\mathbf{M}_{a_k}} = \widetilde{\mathbf{V}}_k\left(\widehat{\mathbf{R}}_k + \mathbf{I} \right)
\end{equation}
and
\begin{equation}
\widehat{\mathbf{D}_{a_k}} = \left(\widetilde{\boldsymbol{\Sigma}}_k + \epsilon_k \mathbf{I}\right)^{-1/2}
\end{equation}
}
where 
\begin{equation}
\epsilon_k = \widehat{\gamma}_k^2 + 2\widehat{\gamma}_k\sigma_{\text{max}}\left(\widetilde{\mathbf{H}}_k^H\widetilde{\mathbf{H}}_k\right).
\end{equation}
The rest of the matrices are defined in Appendix A and not included here for the sake of brevity.
\begin{proof}
See Appendix A.
\end{proof}

Remarkably, whenever $\epsilon_k$ increases, the resulting robust precoding design is more different than the original design $\mathbf{W}_{a_k}^{\text{MMSE}}$. In any case, the computational complexity of the robust design remains the same.

\subsection{Robust Intra-beam Precoding}

Similarly to the previous robust design the intra-beam precoding shall consider tentative perturbations on their channel matrices. Considering the average optimization scheme, worst-case robust optimization for the $k$-th beam can be described as the following optimization problem
\begin{equation}\label{rmaxmeanwb}
\begin{aligned}
& \underset{\mathbf{w}_{b_k}}{\text{maximize}} \quad \underset{\boldsymbol{\Delta}_k}{\text{minimize}}  \quad \|\widehat{\mathbf{Z}}_{k}\mathbf{w}_{b_k}\|^2 \\
& \text{subject to}\\
&\|\mathbf{w}_{b_k}\|^2 = 1,  \\
&\|\boldsymbol{\Delta}_k\|^2 \geq \underline{\gamma_k},  \\
\end{aligned}
\end{equation}
where 
\begin{equation}
\widehat{\mathbf{Z}}_{k} = \mathbf{H}_k\mathbf{W}_{a_k} + \boldsymbol{\Delta}_k\mathbf{W}_{a_k}.
\end{equation}
The next theorem presents an approximate solution of the aforementioned problem.

\textbf{Theorem 2} \emph{ An approximate solution of \eqref{rmaxmeanwb} is $\widehat{\mathbf{z}}_{k_1}$, which is the first column vector of matrix
\begin{equation}
\mathbf{L}_k\left(\nu_k \mathbf{N} \circ \left(\mathbf{L}_k^H\mathbf{L}_k^H\mathbf{T}_k + \mathbf{T}_k\mathbf{L}_k^H\mathbf{L}_k\right) + \mathbf{I} \right)
\end{equation}
where 
\begin{equation}
\nu_k = \underline{\gamma}_k\mathbf{1}^T\left(\widetilde{\boldsymbol{\Sigma}}_k + \epsilon_k \mathbf{I}\right)^{-1/2}\mathbf{1},
\label{kakh3}
\end{equation}
and the rest of matrices are defined in Appendix B.}
\begin{proof}
See Appendix B.
\end{proof}
Note that whenever $\nu_k$ increases, the solutions is more different than the original solution. Again, the computational complexity remains the same as the non-robust case.

\section{User Grouping}

One of the main limiting factors of multibeam multicast precoding is the spatial diversity of the users within each beam. Indeed, whenever the targeted users in each beam have orthogonal channel vectors, this is,
\begin{equation}
\mathbf{h}_{k,m}^H\mathbf{h}_{k,n} = 0,
\end{equation}
for the $k$-th beam for $m,n=1 \ldots , Q$ and $m \neq n$; the intra-beam precoding is not able to deliver the intended symbols. Under this context it is beneficial that the system designer performs a user grouping before the precoding matrix is computed so that users with collinear channel vectors are simultaneously served

\subsection{$k$-User Grouping}

In a given time instant, the scheduler determines a set of tentative users to be served. The number of these scheduled users is considered the same for each beam, fixed and equal to $\mathcal{Q}$. For each beam, obtaining the most adequate groups of users is a cumbersome problem. Note that first, the system designer shall determine the adequate number of groups $G_k$ per beam and, posteriorly, group them into those $G_k$ groups. Clearly, the overall system throughput will depend on the user density over the coverage area: the larger number of users over the coverage, the larger throughputs can be obtained. 

In order to solve this problem, we will consider a random pre-processing. This pre-processing consists of randomly choose a user from the beam and, posteriorly, obtain the group of users. Note that with this first processing, we are severely levering the computational complexity of the technique.

Under this context, let us consider that we have elected an arbitrary user $m$ within the $k$-th beam. The user grouping scheme shall obtain the closest $Q-1$ users in terms of Euclidean norm. Mathematically,
\begin{equation}\label{group}
\begin{aligned}
& \underset{n \in {1, \ldots , Q}}{\text{minimize}}  \quad \| \mathbf{h}_{k,m}  -  \mathbf{h}_{k,n}\|^2 \\
& \text{subject to}\\
& n \neq m.  \\
\end{aligned}
\end{equation}

Since the considered $Q$ in \eqref{group} is not expected to be large, this optimization only requires a set of Euclidean norm comparisons between the scheduled users. As it happens with the precoding design, the user grouping scheme suffers from degradation whenever the user channel vectors are corrupted. A method for robustly group them is presented in the next subsection.

\subsection{Robust $k$-User Grouping}

Whenever the channel vectors are corrupted by a certain perturbation, a worst-case optimization shall be performed

\begin{equation}\label{rgroup}
\begin{aligned}
& \underset{n \in {1, \ldots , Q}}{\text{minimize}} \quad \underset{\{\boldsymbol{\delta}_{k,q}\}_{q = 1}^Q}{\text{maximize}}  \quad \| \mathbf{h}_{k,m} + \boldsymbol{\delta}_{k,m}  -  \mathbf{h}_{k,n} - \boldsymbol{\delta}_{k,n}\|^2 \\
& \text{subject to}\\
& n \neq m,  \\
& \|\boldsymbol{\delta}_{k,q}\|^2 \leq  \gamma_{k,q}.\\
\end{aligned}
\end{equation}

Next theorem provides an approximate version of the aforementioned problem.

\textbf{Theorem 3} \emph{ An optimization problem whose solutions upper bound the original  worst-case robust grouping problem \eqref{rgroup} is}

\begin{equation}\label{rgroup2}
\begin{aligned}
& \underset{n \in {1, \ldots , Q}}{\text{minimize}}   \quad \| \mathbf{h}_{k,m}   -  \mathbf{h}_{k,n} \|^2 + \gamma_{n,q} \\
& \text{subject to}\\
& n \neq m.  \\
\end{aligned}
\end{equation}

\begin{proof}
It is a simple derivation considering the Cauchy-Schwarz inequality and the fact that given a randomly chosen user $m$ its perturbation does not influence the grouping optimization.
\end{proof}

With this optimization it is evident that whenever uncertainty is assumed in the channel vectors, this shall be considered in the user grouping design by means of an additional scalar penalty. Remarkably, in case we consider the same uncertainty to all users, the proposed approximate solution remains the same.

\section{Multiple Gateway Architecture}

As a matter of fact, there might be the case where the feeder link cannot support the overall satellite data traffic. For instance, a payload equipped with $N = 155$ feeds with a user channel bandwidth of 500 MHz requires a feeder link bandwidth of 77.5 GHz which is an unaffordable requirement even if the feeder link carrier is located at the Q/V band.

\begin{figure}[h!]
  \vspace{-1.3 cm}
	\centering
    \includegraphics[width=14.5cm, height= 10cm]{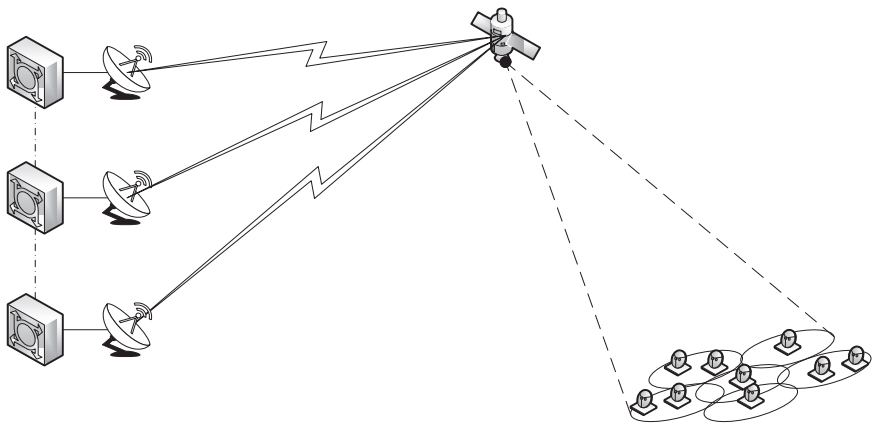}
     \vspace{-2.7 cm}
		\caption{The picture depicts the multiple gateway multicast multibeam satellite structure. In contrast to the single gateway architecture, several gateways transmit the data to be delivered to the coverage area. This entails two main drawbacks. First, certain inter-gateway link (dotted lines) shall exist. Second, each ground unit must compute an independent submatrix of the overall precoding matrix.}
      \label{systemMG}
\end{figure}

In order to solve this problem, the feeder link might benefit from certain spatial reuse so that several gateways can simultaneously send the data to be transmitted over the satellite coverage area. Under this context, $G$ gateways can reuse the available bandwidth for the feeder link leading to large increase of the user bandwidth (see Figure \ref{systemMG}). However, in a multiple gateway scenario, the precoding scheme shall be reconsidered.

In order to keep the payload complexity low, each feeder link receiver at the payload will only route signals a set of feeds $N_g$ for $g = 1, \ldots , G$. Otherwise, a very complex analog scheme shall be implemented. In addition, it is important to remark that the feeder link bandwidth resources can only support a small number of $N_g$ feed signals. Considering this approach, the precoding matrix $\mathbf{W}$ must be partitioned into $G$ sub-matrices leading to a large decrease of the achievable throughput. In the following subsections we propose some techniques in order to overcome this limitation.

In addition, considering that in order to reduce the inter-gateway communication, each gateway individually computes its precoding sub-matrix, certain cooperative scheme shall be conceived. Indeed, each gateway only have access to the feedback from its corresponding set of feeds $N_g$. With this, certain CSI cooperation among gateways shall be established so that inter-beam interference is mitigated.

\subsection{Precoding Scheme}

As discussed in the previous section, the $g$-th  gateway only has access to a set of $N_g$ of the overall feed elements located at the payload. Additionally, the gateway serves a set of $K_g$ beams out of the $K$, leading to a total amount of served users equal to $K_gQ$.  We will assume a known feed allocation per gateway and fixed $N_g$ and $K_g$. Precisely, we will consider that the feeds are assigned in a consecutive fashion over the channel matrix as it is defined in the following.

Let us assume that each gateway has access to the overall CSI. With this, the matrix to be transmitted through the feeder link of the $g$-th gateway jointly with the user symbols is
\begin{equation}
\mathbf{W}^g \in \mathbb{C}^{N_g \times K_g}.
\end{equation}
In other words, as each gateway only have access to a certain set of feed elements the symbols to be transmitted to the $K_g$ beams shall be linearly transformed with $\mathbf{W}^g$ in order to increase the overall throughput. Whenever each gateway has access to the overall channel matrix $\mathbf{H}$, the precoding scheme can be computed as described in Section III and adapted to the multiple gateway scenario. This can be done by beams of setting zero entries in the precoding matrix whenever the gateway does not have access to a certain feed.  This transformation leads to a block diagonal precoding matrix as follows
\begin{equation}
\mathbf{W}_{\text{M-GW}} = 
\begin{bmatrix}
  \mathbf{W}^1 & \mathbf{0} & \cdots & \mathbf{0} \\
  \mathbf{0} & \mathbf{W}^2 & \cdots & \mathbf{0} \\
  \vdots  & \vdots  & \ddots & \vdots  \\
  \mathbf{0} & \mathbf{0} & \cdots & \mathbf{W}^G
 \end{bmatrix}
 \in \mathbb{C}^{N \times K}.
\end{equation}
Evidently, robust designs can be applied without any additional penalty. 

\subsection{CSIT Sharing}

Since the gateway only has access to certain beams, its available CSI is reduced  Precisely, the $g$-th  receives from its feedback link the following matrix
\begin{equation}
\mathbf{H}^g = \mathbf{H}\left( \left(\left( (g-1)QK_g + 1 \right):gQK_g,  1:N  \right) \right) \in \mathbb{C}^{QK_g \times N_g},
\end{equation}
where the Matlab notation has been used for the sake of clarity. However, in order to compute the precoding matrix, the gateways need the channel effect between their assigned feeds to the non-intended users.  With this, every gateway must transmit over the inter-gateway link (e.g. a fiber optic) the information fed back from its users corresponding the effect of the non-assigned feeds. Mathematically, the $g$-th gateway must share
\begin{equation}
 \left( \mathbf{H}^{g_l} \right)_{l\neq g}^G,
\end{equation}
where 
\begin{equation}
\mathbf{H}^{g_l} = \mathbf{H}\left( \left(  \left((g-1)QK_g + 1 \right):gQK_g \right), ((l-1)N_g + 1):gN_g \right).
\end{equation}
This cooperation among gateways require a total amount of
\begin{equation}
(G-1)QK_gN_g
\end{equation}
complex numbers to be shared by each gateway. As a consequence, it is essential to reduce this communication overhead in order to reduce the overall system cost. 

One approach is to limit the sharing between the different gateways and only consider the $C$ closer gateways. With this, the overall data overhead reduces to
\begin{equation}
CQK_gN_g.
\end{equation}
Another alternative is to apply certain compression to the transmit channel submatrices. For instance, we could use the eigenvector associated to the largest eigenvalue of each matrix $\mathbf{H}^{g_l}$. This will lead to a total communication overhead of
\begin{equation}
(G-1)N_g.
\end{equation}

Evidently, whenever each gateway has a more precise version of the channel matrix, the larger throughput can be obtained. In the simulation section this is evaluated and different cooperation schemes are evaluated.

\section{Simulations}

\begin{table}[h!]
\caption{USER LINK SIMULATION PARAMETERS} 
\begin{center} 
\begin{tabular}{ | l | l |}
\hline
\begin{math}\textbf{Parameter}\end{math}&\begin{math}\textbf{Value}\end{math} \\ \hline
Satellite height & 35786 km (geostationary) \\ \hline
Satellite longitude, latitude &\begin{math} 10^\circ East, 0^\circ\end{math} \\ \hline
Earth radius& 6378.137 Km   \\ \hline
Feed radiation pattern & Provided by\begin{math} ESA\end{math} \\ \hline
Number of feeds N & 245  \\ \hline
Beamforming matrix $\textbf{B}$&Provided by ESA\\ \hline
Number of beams&100\\ \hline
User location distribution&Uniformly distributed \\ \hline
Carrier frequency&20 GHz (Ka band) \\ \hline
Total bandwidth&500 MHz \\ \hline
Roll-off factor&0.25 \\ \hline
User antenna gain&41.7 dBi \\ \hline
$G/T$  & 17.68 dB/K \\ \hline
\end{tabular}
     \vspace{-0.6cm}  \label{tab:label}
\end{center}
\end{table}
Considering a reference scenario of a geostationary satellite with $N=K=245$, we evaluate the proposed precoding designs considering a full frequency reuse scenario. Array fed radiation pattern has been provided by the European Space Agency and it takes into account the different user locations over the European continent. The link budget parameters are described in Table 1. 

All results have been obtained considering 500 channel realizations and a phase effect between feeds $\chi = 10º$ degrees. Moreover, throughput values are obtained by means of the user SINR and the efficiency (bit/symbol) given a minimum Packet Error Rate (PER) of 10$^{-6}$ considering DVB-S2X. It is important to remark that this relationship has been obtained from \cite{ETSI} considering the PER curves.

The outline of the subsequent subsections is as follows. First, we show the performance gain of the proposed precoding schemes considering perfect CSI and single gateway architecture. Second, it is shown that larger throughputs can be obtained if user grouping techniques are applied. Third, the impact of imperfect CSI is shown and the convenience of robust designs is presented. Finally, the multiple gateway architecture is evaluated so as the proposed inter-gateway cooperation techniques. Remarkably, for a best practice we also consider a reference scenario that consists in 4-colouring scheme where the interference is reduced so as the available bandwidth.

%

\begin{figure*}[tb!]
\centering \small \subfigure 
    {%
       \hspace*{-2cm}\includegraphics[width=9.5cm,height=7cm]{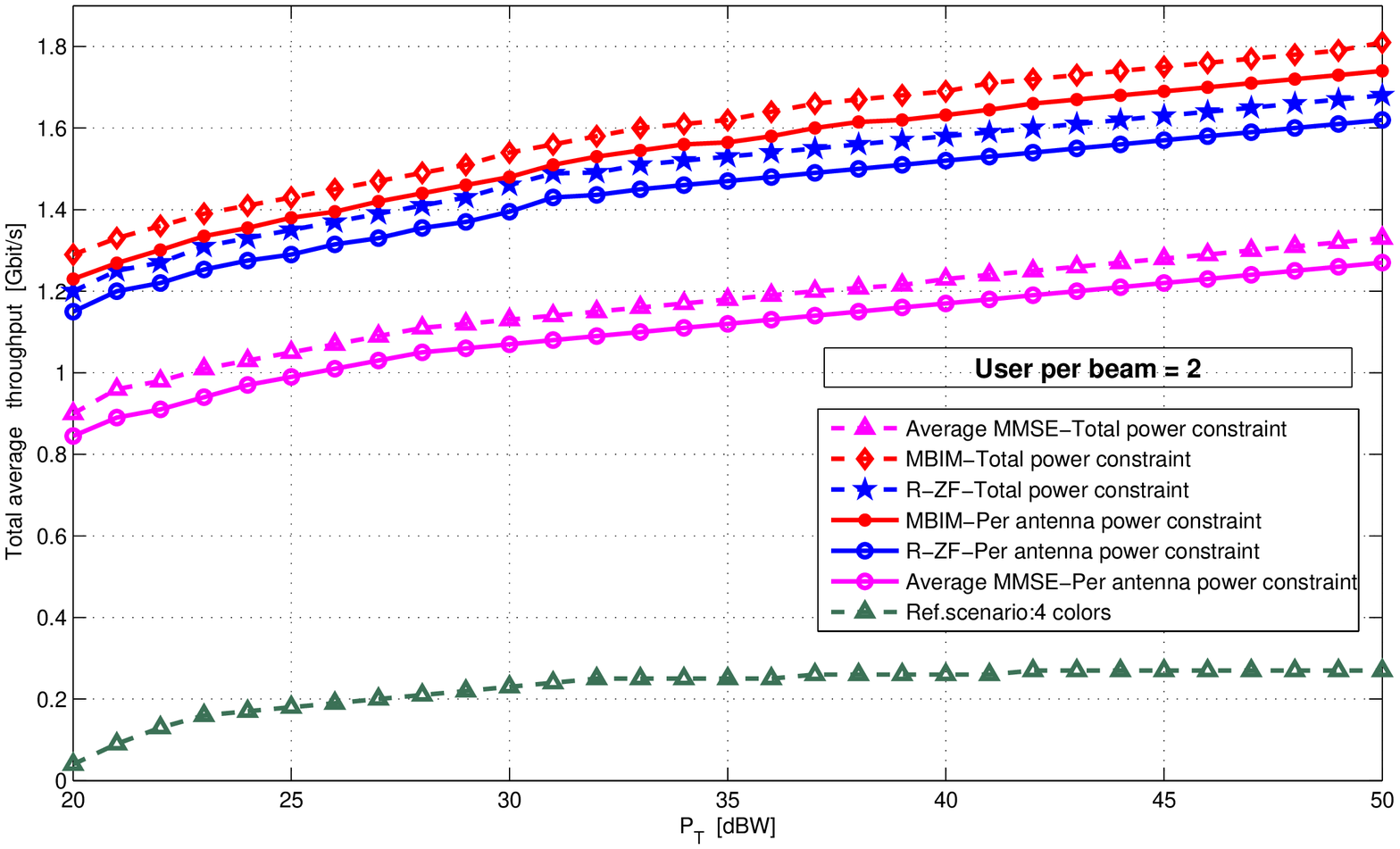}
 		}%
\subfigure
 		{%
       \hspace*{-0.5cm} \includegraphics[width=9.5cm,height=7cm]{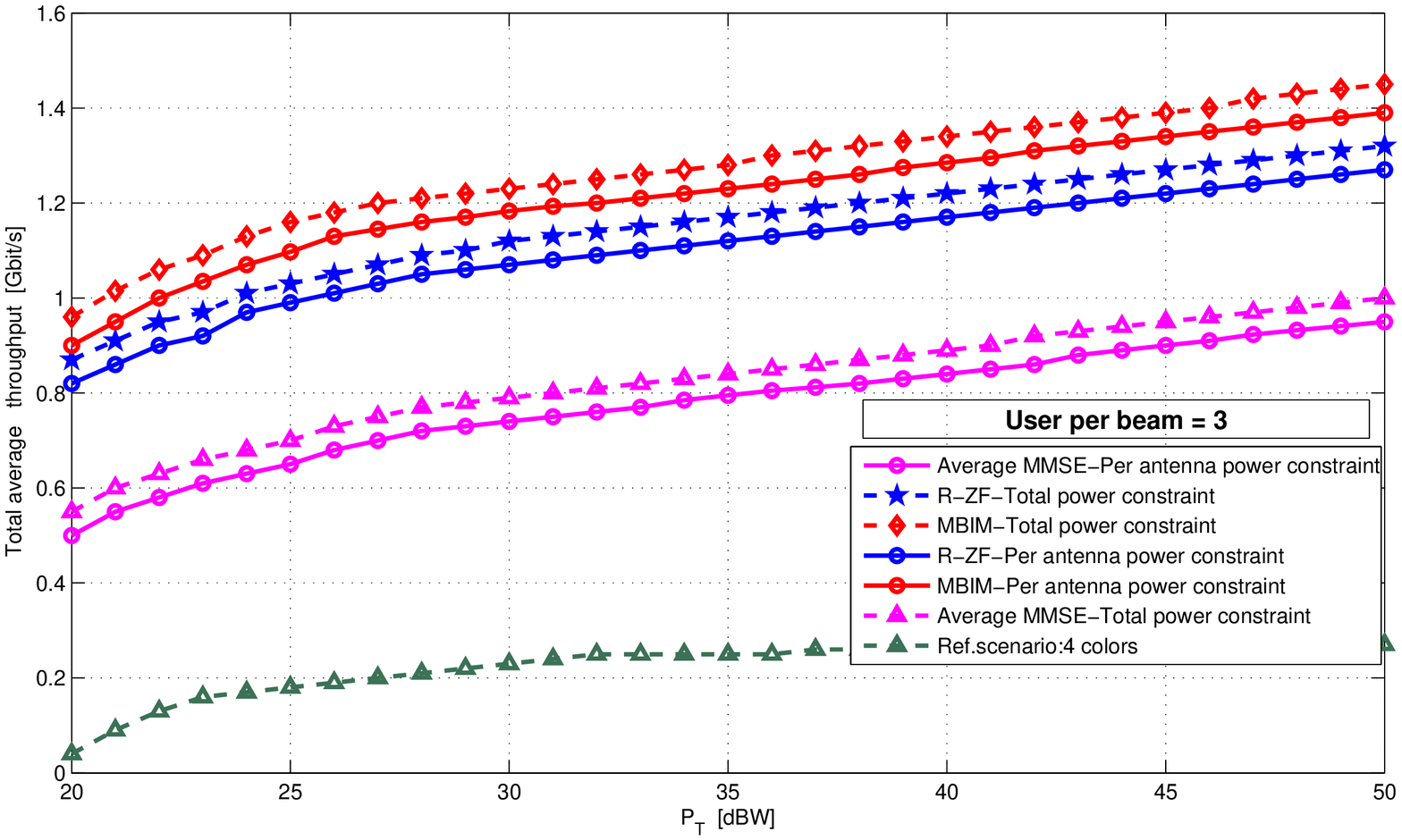}
    }%
    \vspace{-0.4cm}\caption{Per beam average  throughput with $Q=2,3$ users per beam, perfect CSI, no user grouping.} \label{Q2} \vspace{-0.6cm}
\end{figure*}

Figure \ref{Q2} presents the system throughput considering the proposed precoding schemes in section III. Both of them, R-ZF and MBIM are compared to the average MMSE 
design presented in \cite{6843054,6934558}. It can be observed that both proposals behave better than average MMSE scheme. Specially, MBIM offers larger throughputs than R-ZF over the different transmit power values. Moreover, the conventional 4-coloring scheme has the lowest achievable rate due to the bandwidth limitations.

%
%

\subsection{User Grouping}

\begin{figure*}[tb!]
\centering \small \subfigure 
    {%
       \hspace*{-2cm}\includegraphics[width=9.5cm,height=7cm]{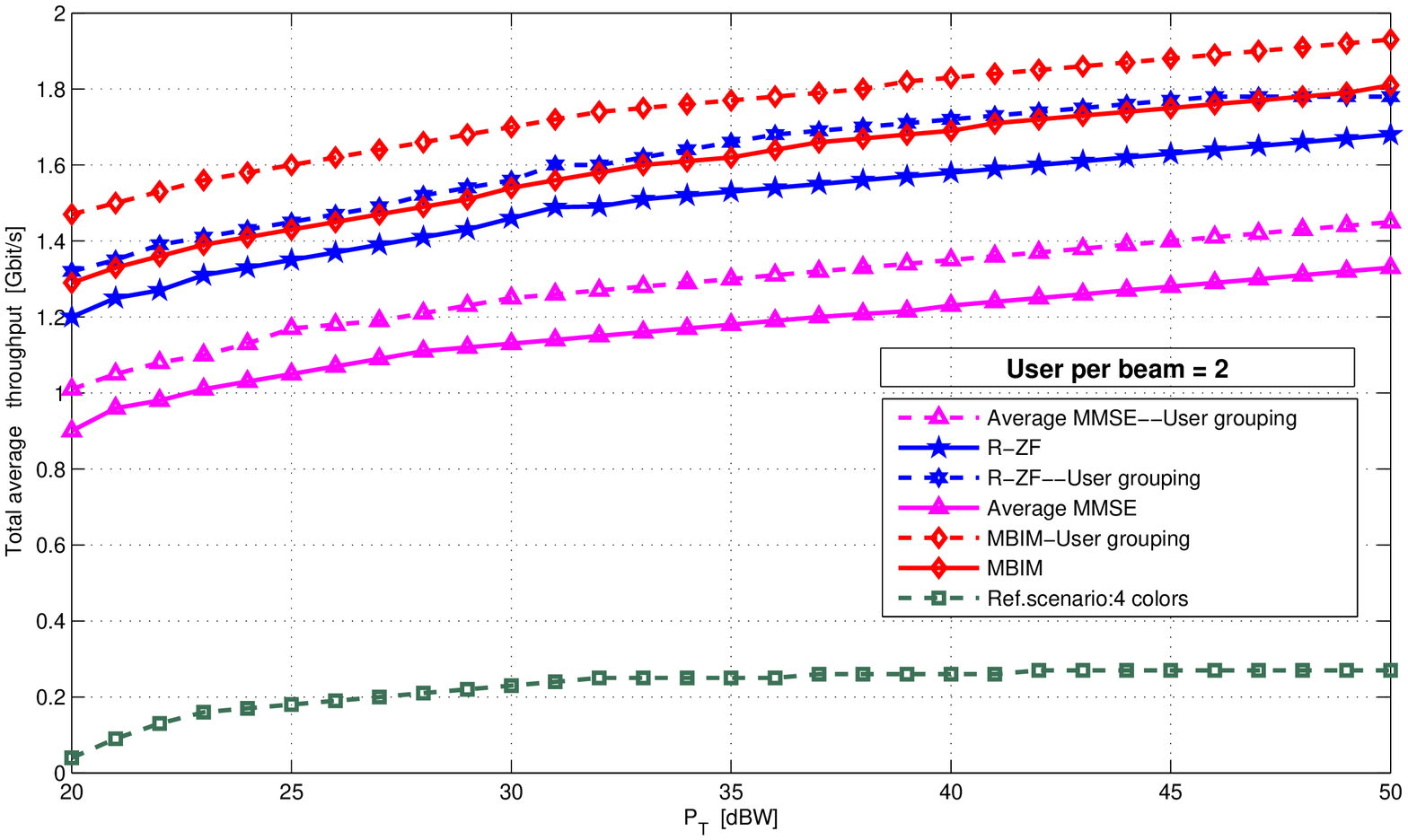}
 		}%
\subfigure
 		{%
       \hspace*{-0.5cm} \includegraphics[width=9.5cm,height=7cm]{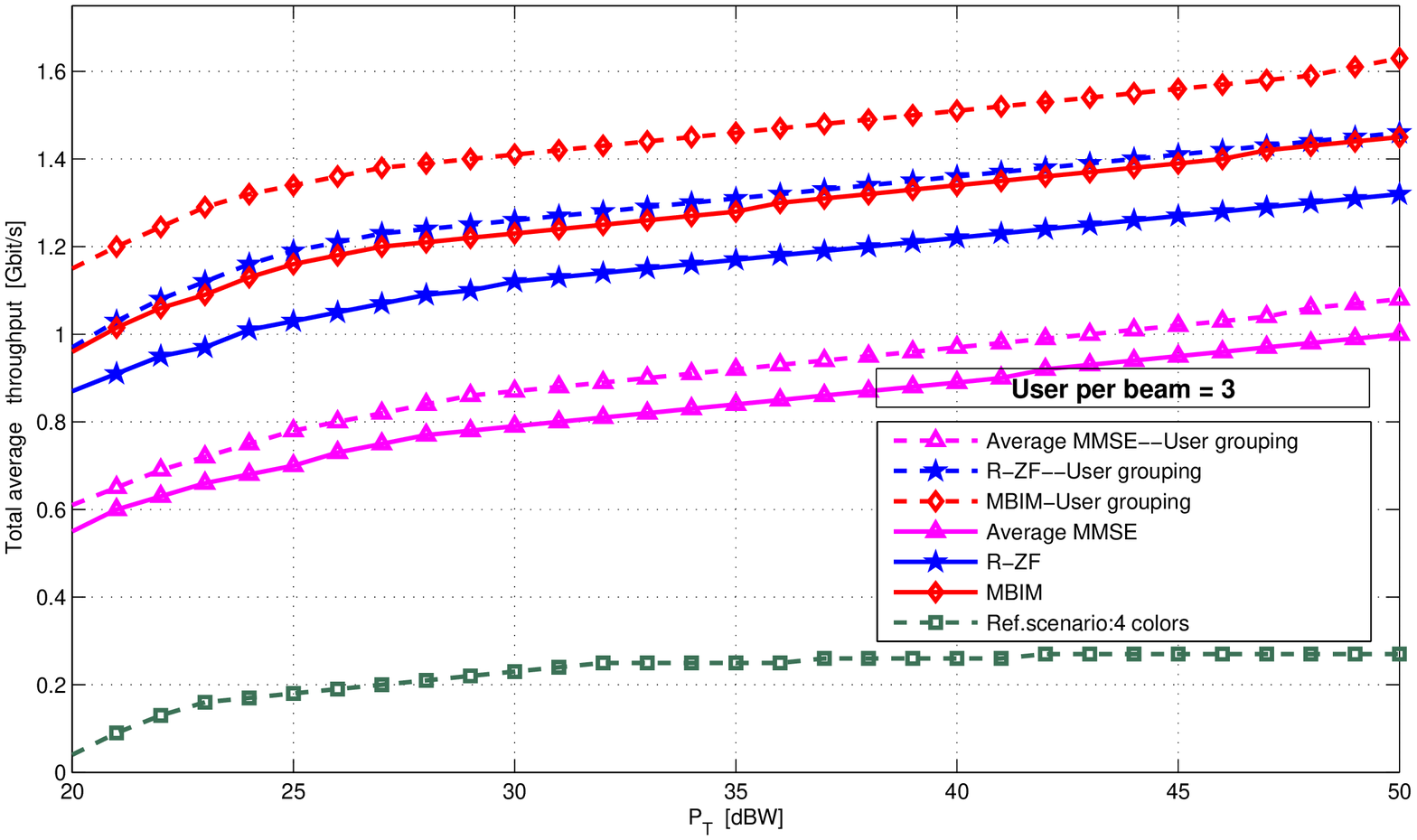}
    }%
    \vspace{-0.4cm}\caption{Per beam average  throughput  with $Q=2,3$ users, perfect CSI, user grouping.} \label{gQ2} \vspace{-0.6cm}
\end{figure*}

As discussed before, whenever the users within the same beam have collinear channel vectors, large rates can be obtained. This is shown in Figure \ref{gQ2}. Similar to the study in \cite{daniel}, in each beam $\mathcal{Q} = 200$ users are uniformly distributed, resulting in an average user density of 0.023 users/$km^{2}$ inside the 3 dB coverage edge of every beam.

It results that in all cases, user grouping increases the throughput. Specifically, for $Q=3$, the user grouping gain becomes larger than in the $Q=2$ case. Considering these results and the associated low complexity operation, it results convenient to implement this technique in satellite systems.

\subsection{Robust Design}

This subsection considers the effect of imperfect CSI at the transmitter. This is modelled with a additive perturbation matrix whose entries are Gaussian distributed with zero mean and variance equal to 1. The perturbation values $\gamma_k$ are considered the same for all submatrices ($K=1, \ldots , K$) and in the simulations are obtained considering that each submatrix does no exceed the assumed bound. Additionally, $\underline{\gamma_k}$ is set to 1 and no sensitivity analysis is performed due to space limitations.
Figure \ref{rQ2}  shows the performance of our proposal for different $\gamma$ values represented in a ratio basis with respect to the channel matrix Frobenious norm.  For both cases $Q=2,3$, the proposed technique is able to overcome the imperfect CSI values at the transmitter leading large throughputs while keep the precoding complexity low.

\subsection{Multiple Gateway Transmission}

In order to evaluate the performance of the proposed multiple gateway schemes, we consider the following methods and inter-gateway architectures. We will consider a total number of $G = 14$ gateways each of them serving either 17 or 18 beams.
\begin{itemize}
%
\item \textbf{Scenario 1 :} Individual cluster processing. Each gateway processes its set of beams independently and
only receives the CSI from its corresponding beams. With this, it is not possible to mitigate the interference of adjacent beams. This is referred to Individual Cluster Processing (ICP).


\item \textbf{Scenario 2 :} Gateway $g$ (respectively for all the gateways) collaborates with 4 gateways that serve beams directly adjacent to their beams so that $g$-th gateway receives prefect CSI of adjacent beams. This is referred to 4 Gateway Collaboration Processing (4GCP). 

\item \textbf{Senario 3 :}Gateway $g$ (respectively for all the gateways) collaborates with 7 gateways that serve beams directly adjacent to their beams so that $g$-th gateway receives prefect CSI of adjacent beams. This is referred to 7 Gateway Collaboration Processing (7GCP). 

\item \textbf{Senario 4 :} Gateway $g$ (respectively for all the gateways) collaborates with all gateways by means of sharing the singular left vector associated with the largest singular value of the  gateway channel matrix. This is referred to Maximum SVD of Gateway Collaboration (MSVDGC).
\begin{figure*}[tb!]
\centering \small \subfigure 
    {%
       \hspace*{-2cm}\includegraphics[width=9.5cm,height=7cm]{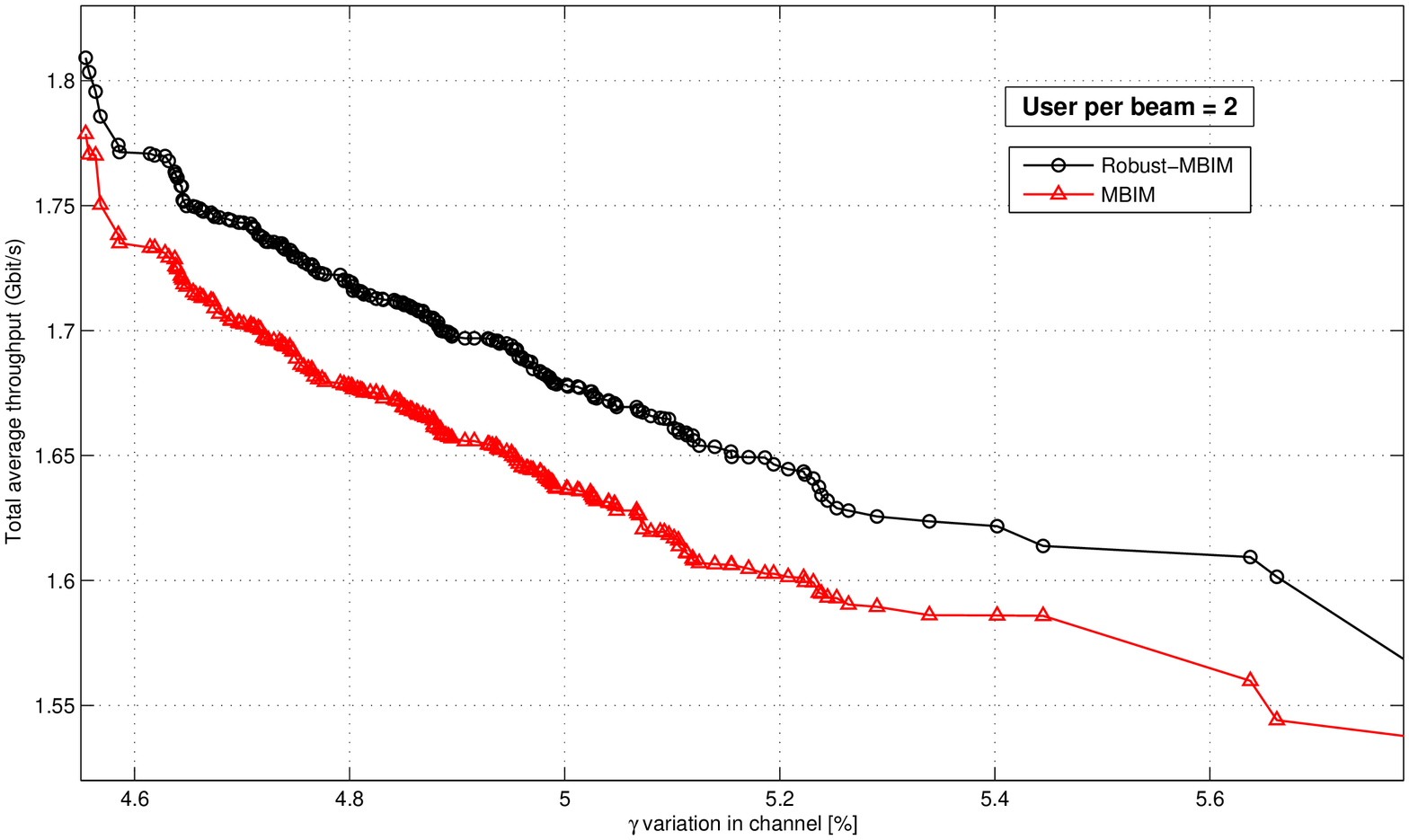}
 		}%
\subfigure
 		{%
       \hspace*{-0.5cm} \includegraphics[width=9.5cm,height=7cm]{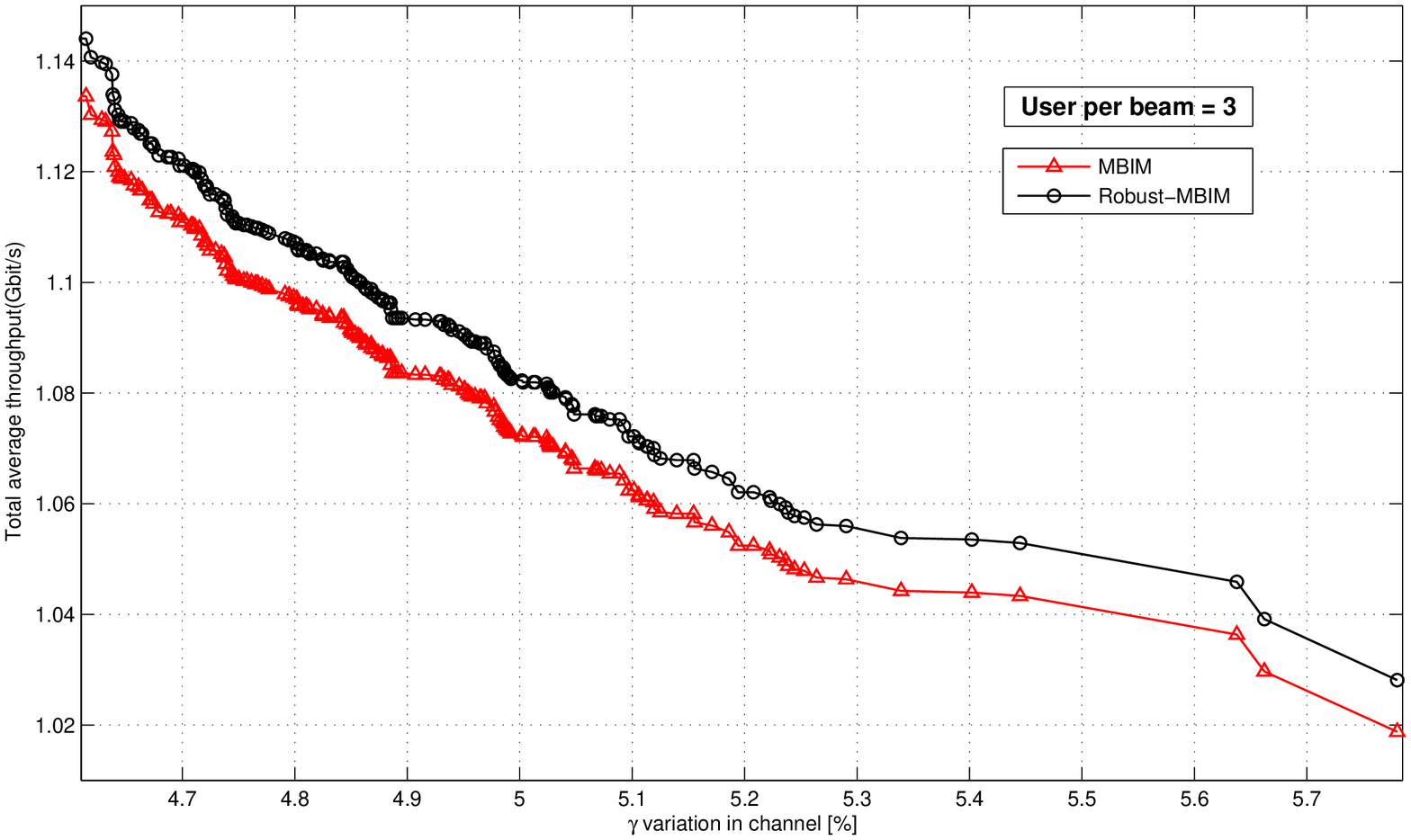}
    }%
    \vspace{-0.4cm}\caption{Per beam average  throughput  with $Q=2,3$ users, imperfect CSI, no user grouping.} \label{rQ2} \vspace{-0.6cm}
\end{figure*}
\item \textbf{Senario 5 :} Single gateway scenario so that a unique on ground processing unit is able to use all available feeds with the overall channel matrix. This scenario refers
to Reference scenario (Ref).

\end{itemize}

\begin{figure*}[tb!]
\centering \small \subfigure 
    {%
       \hspace*{-2cm}\includegraphics[width=9.5cm,height=7cm]{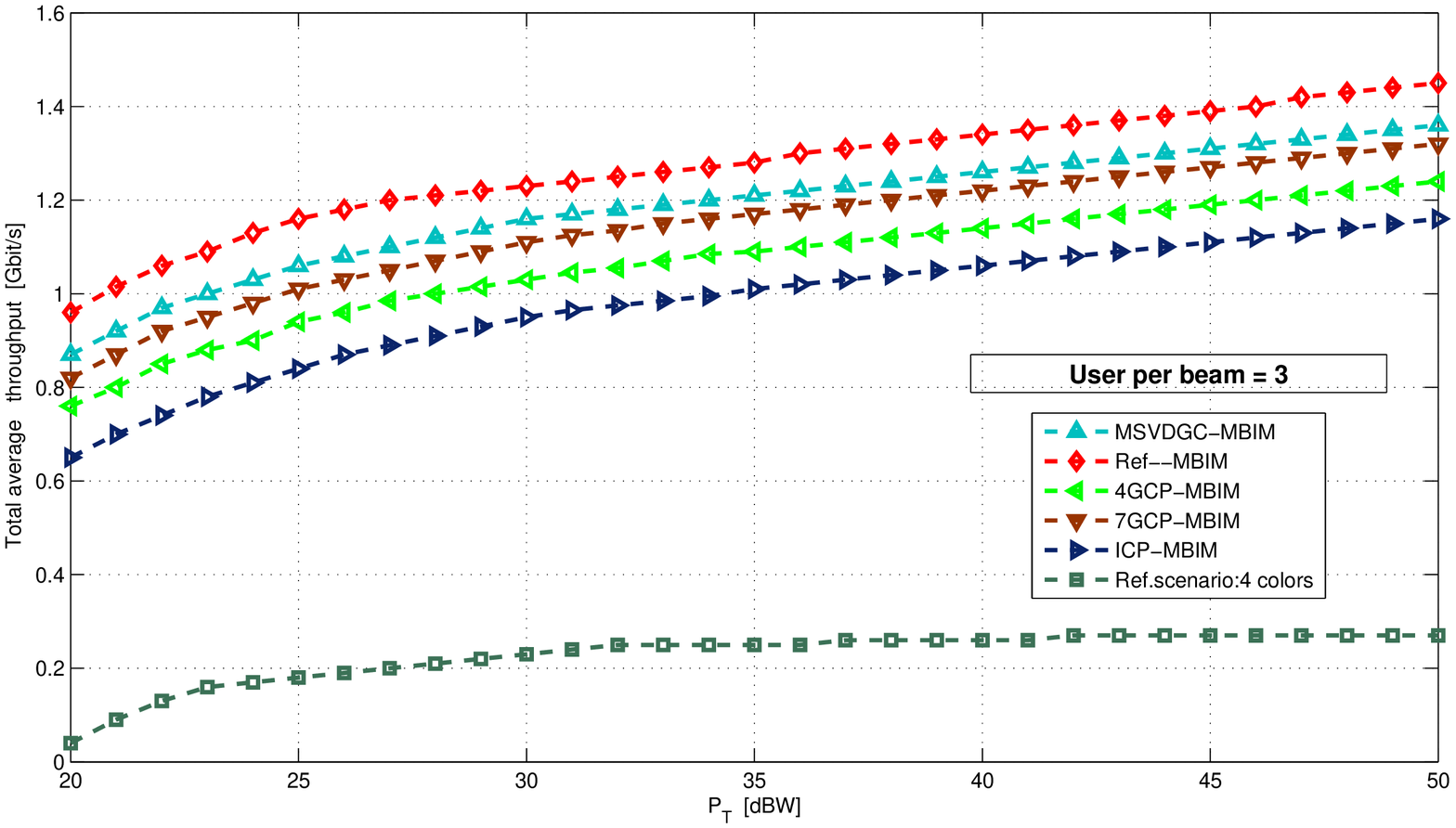}
 		}%
\subfigure
 		{%
       \hspace*{-0.5cm} \includegraphics[width=9.5cm,height=7cm]{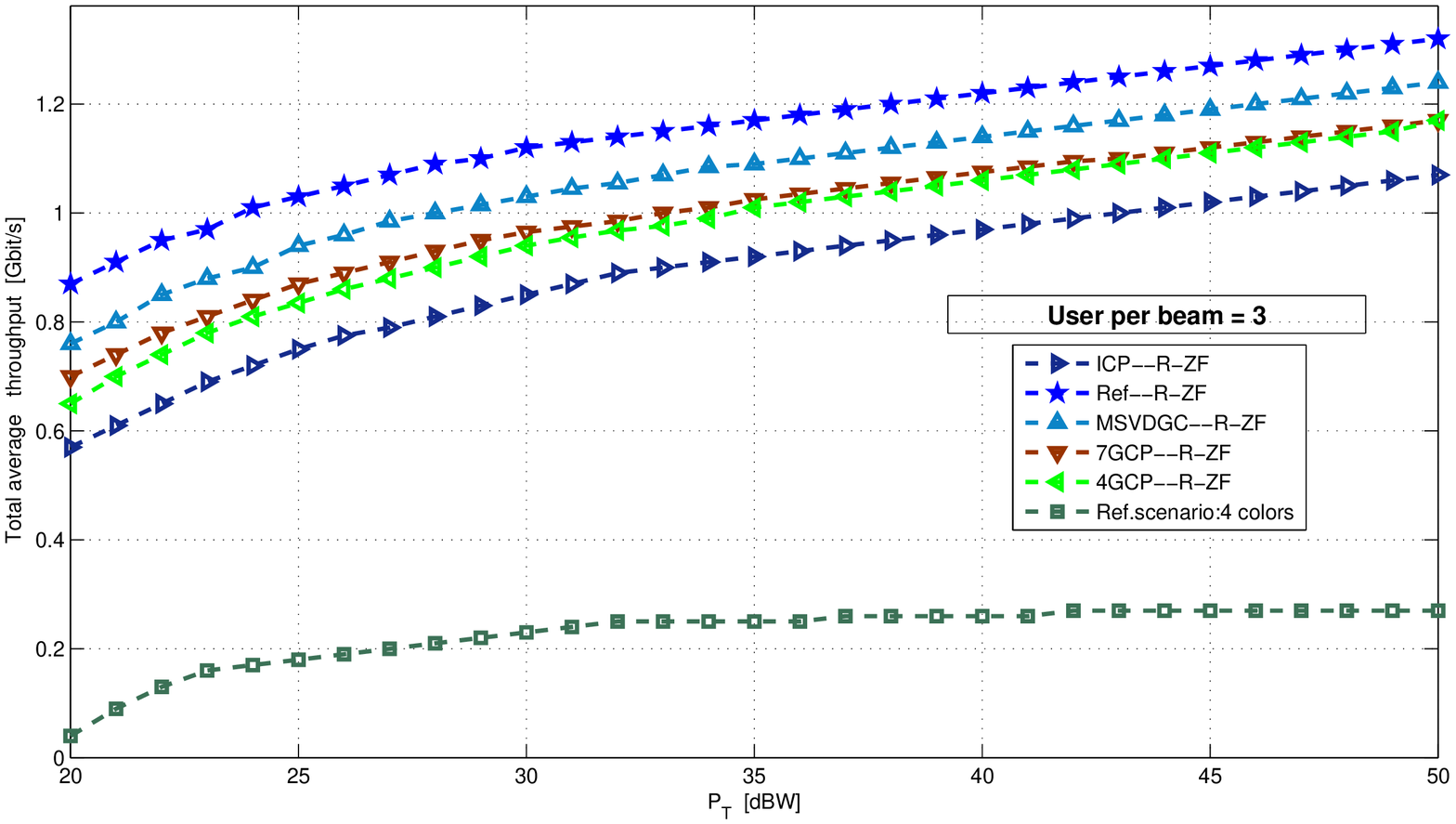}
    }%
    \vspace{-0.4cm}\caption{Per beam average  throughput  with $Q=3$ users, perfect CSI, no user grouping and different multiple gateway architectures with MBIR precoding (Left), R-ZF precoding(Right).} \label{MBIRQ3} \vspace{-0.6cm}
\end{figure*}

Figure \ref{MBIRQ3}  shows the proposed multiple gateway architectures for both the R-ZF and MBIR precoding designs. Evidently, as it happened in the single gateway scenario, MBIR provides larger overall throughputs than the R-ZF in all cases. In addition, it is observed that the larger cooperation is considered, the larger throughputs are obtained. Remarkably, MSVDGC method offers a good trade-off between overall throughput and inter-gateway communication overhead.

\section{Conclusions}

This paper proposed a two-stage low complex precoding design for multibeam multicast satellite systems. While the first stage minimizes the inter-beam interference, the 
%
second stage enhances the intra-beam SINR. A robust version of the proposed scheme is provided based on the novel approach of the first perturbation theory. Additionally, user grouping and multiple gateway schemes are presented as essential tools for increasing the throughput in multibeam satellite systems. The conceived methods are evaluated in a continental coverage area and they result to perform better than the current approaches yet offering a low computational complexity.

\appendices
\section{Proof of Theorem 1}

After some manipulations, the objective function can be written
\begin{equation}
\sum_{k=1}^K \text{trace} \left(\mathbf{W}_{a_k}^H\left(\widehat{\widetilde{\mathbf{H}}}_k^H\widehat{\widetilde{\mathbf{H}}}_k + \frac{KQ}{P_{T}}\mathbf{I} \right)\mathbf{W}_{a_k} \right).
\end{equation}
This objective function can be re-written considering the eigendecomposition of $\widehat{\widetilde{\mathbf{H}}}_k\widehat{\widetilde{\mathbf{H}}}_k^H = \widehat{\widetilde{\mathbf{V}}}_k\widehat{\widetilde{\boldsymbol{\Sigma}}}_k\widehat{\widetilde{\mathbf{V}}}_k^H $, so that
\begin{equation}\label{objective}
\sum_{k=1}^K \text{trace} \left(\mathbf{W}_{a_k}^H\widehat{\widetilde{\mathbf{V}}}_k\left(\widehat{\widetilde{\boldsymbol{\Sigma}}}_k + \frac{KQ}{P_{T}}\mathbf{I} \right)\widehat{\widetilde{\mathbf{V}}}_k^H\mathbf{W}_{a_k} \right).
\end{equation}
Since the objective function depends on the perturbation matrices and they are unknown by the transmitter, the following derivations aim at obtaining an upper-bound of \eqref{objective}. Concretely, an upper-bound of both $\widehat{\widetilde{\mathbf{V}}}_k$ and $\widehat{\widetilde{\boldsymbol{\Sigma}}}_k$ will be presented. The optimization problem \eqref{mmse} can yield to a tractable solution.

Proceeding with the derivation, the following equality holds
\begin{equation}
\widehat{\widetilde{\mathbf{H}}}_k^H\widehat{\widetilde{\mathbf{H}}}_k = \widetilde{\mathbf{H}}_k^H\widetilde{\mathbf{H}}_k
+ \widetilde{\boldsymbol{\Delta}}_k^H\widetilde{\mathbf{H}}_k + \widetilde{\mathbf{H}}_k^H\widetilde{\boldsymbol{\Delta}}_k + \widetilde{\boldsymbol{\Delta}}_k^H\widetilde{\boldsymbol{\Delta}}_k,
\end{equation}
where it can be observed that the first term is the exact term whereas the rest are perturbation terms. In the following, we will consider the perturbation effect of 
\begin{equation}
\Delta\mathbf{K}_k = \widetilde{\boldsymbol{\Delta}}_k^H\widetilde{\mathbf{H}}_k + \widetilde{\mathbf{H}}_k^H\widetilde{\boldsymbol{\Delta}}_k + \widetilde{\boldsymbol{\Delta}}_k^H\widetilde{\boldsymbol{\Delta}}_k,
\end{equation}
in the precoding design.

Considering the first order perturbation analysis presented in \cite{Liu08}, the eigenvectors of $\widehat{\widetilde{\mathbf{H}}}_k^H\widehat{\widetilde{\mathbf{H}}}_k$ can be written as
\begin{equation}
\widehat{\widetilde{\mathbf{V}}}_k = \widetilde{\mathbf{V}}_k\left(\mathbf{R}_k + \mathbf{I} \right),
\end{equation}
where 
\begin{equation}
\mathbf{R}_k = \mathbf{D} \circ \left(\widetilde{\mathbf{V}}_k^H\Delta\mathbf{K}\widetilde{\mathbf{V}}_k^H\widetilde{\boldsymbol{\Sigma}}_k + \widetilde{\boldsymbol{\Sigma}}_k\widetilde{\mathbf{V}}_k^H \Delta\mathbf{K}^H\widetilde{\mathbf{V}}_k\right).
\end{equation}
where the $g$,$f$-entry of $\mathbf{D}$ is
\begin{equation}
\frac{1}{\lambda_f - \lambda_g},
\end{equation}
for $f \neq g$ and 0 whenever $f=g$. $\lambda_f$ denotes the $f$-th eigenvalue of $\widetilde{\mathbf{H}}_k^H\widetilde{\mathbf{H}}_k$. It is important to remark that it has been considered that $\widetilde{\mathbf{H}}_k^H\widetilde{\mathbf{H}}_k$ is full rank and, therefore, its null space has 0 dimension.

From the problem definition $\eqref{mmse}$ it is possible to bound $\Delta\mathbf{K}$ such that
\begin{equation}\label{bound}
\Delta\mathbf{K}_k \preceq \left(\widehat{\gamma}_k^2 + 2\widehat{\gamma}_k\sigma_{\text{max}}\left(\widetilde{\mathbf{H}}_k^H\widetilde{\mathbf{H}}_k\right) \right) \mathbf{I}.
\end{equation} 
With this and considering the following lemma:

\textbf{Lemma 1}. \emph{For any semidefinite positive square matrices $\mathbf{A}, \mathbf{B}, \mathbf{C}$ and $\mathbf{B} \preceq \mathbf{C}$ it holds that 
\begin{equation}
\mathbf{A} \circ \mathbf{B} \preceq \mathbf{A} \circ \mathbf{C}
\end{equation}
}
\begin{proof}
See Theorem 17 of \cite{ham78}.
\end{proof}

It is possible to write the following inequality
\begin{equation}\label{vectors}
\widehat{\widetilde{\mathbf{V}}}_k \preceq \widetilde{\mathbf{V}}_k\left(\widehat{\mathbf{R}}_k + \mathbf{I} \right),
\end{equation}
where
\begin{equation}
\widehat{\mathbf{R}}_k = \epsilon_k \mathbf{D} \circ \left(\widetilde{\mathbf{V}}_k^H\widetilde{\mathbf{V}}_k^H\widetilde{\boldsymbol{\Sigma}}_k + \widetilde{\boldsymbol{\Sigma}}_k\widetilde{\mathbf{V}}_k^H \widetilde{\mathbf{V}}_k\right),
\end{equation}
where 
\begin{equation}
\epsilon = \widehat{\gamma}_k^2 + 2\widehat{\gamma}_k\sigma_{\text{max}}\left(\widetilde{\mathbf{H}}_k^H\widetilde{\mathbf{H}}_k\right)
\end{equation}

In case the perturbations of the eigenvectors are consider, the Weyl's inequality can support the approximation

\textbf{Lemma 2} (Weyl's inequality) Given two Hermitian semidefinite positive matrices $\mathbf{A}_1$ and $\mathbf{A}_2$ with their corresponding eigenvalues collapsed in the diagonal matrices $\boldsymbol{\Sigma}_1$, $\boldsymbol{\Sigma}_2$, the following inequality holds
\begin{equation}
\boldsymbol{\Sigma}_{\text{sum}} \preceq \boldsymbol{\Sigma}_1 + \boldsymbol{\Sigma}_2,
\end{equation}
where $\boldsymbol{\Sigma}_{\text{sum}}$ is a diagonal matrix whose diagonal entries are the eigenvalues of $\mathbf{A}_1 + \mathbf{A}_2$.
\begin{proof}
See \cite{weyl}.
\end{proof}
With this, the eigenvalues of $\widetilde{\mathbf{H}}_k^H\widetilde{\mathbf{H}}_k$ can be upper-bounded so that
\begin{equation}\label{values}
\widehat{\widetilde{\boldsymbol{\Sigma}}}_k \preceq \widetilde{\boldsymbol{\Sigma}}_k + \epsilon_k \mathbf{I}.
\end{equation}

With both \eqref{vectors} and \eqref{values}, it is possible to relax the worst-case maximization in \eqref{rmmse}. In other words, it is possible to use \eqref{vectors} and \eqref{values} as eigendecomposition of matrix $\widehat{\widetilde{\mathbf{H}}}_k$.
\section{Proof of Theorem 2}

As for the previous theorem, the main objective is to find a bound of the objective function so that the dependence with respect to $\boldsymbol{\Delta}_k$ variable disappears.

Considering the first order perturbation model presented in \cite{Liu08}, the required eigendecomposition of $\mathbf{H}_k\mathbf{W}_{a_k}$ is perturbed by the following matrix
\begin{equation}
\Delta\mathbf{Q}_k = \boldsymbol{\Delta}_k\mathbf{W}_{a_k}.
\end{equation}
This leads to the following approximation of the eigenvectors of $\mathbf{Z}_k$
\begin{equation}
\widehat{\mathbf{Z}_k} = \mathbf{L}_k\left(\mathbf{M}_k + \mathbf{I} \right),
\end{equation}
where $\mathbf{L}_k$ are the eigenvectors of $\mathbf{Z}_k$ and 
\begin{equation}
\mathbf{M}_k =  \mathbf{N} \circ \left(\mathbf{L}_k^H\Delta\mathbf{Q}_k\mathbf{L}_k^H\mathbf{T}_k + \mathbf{T}_k\mathbf{L}_k^H\Delta\mathbf{Q}_k^H\mathbf{L}_k \right)
\end{equation}
where $\mathbf{T}_k$ is a diagonal matrix containing the eigenvalues of $\mathbf{Z}_k$ and the $g,f$-th entry of $\mathbf{N}$ is
\begin{equation}
\frac{1}{z_{k_f} - z_{k_g}},
\end{equation}
for $f \neq g$ and 0 whenever $f =g$. $z_{k_f}$ denotes the $f$-th eigenvalue of $\mathbf{Z}_k$.

After some manipulations, we can lower-bound $\Delta\mathbf{Q}_k$ so that
\begin{equation}
\Delta\mathbf{Q}_k \succeq \underline{\gamma}_k\mathbf{1}^T\left(\widetilde{\boldsymbol{\Sigma}}_k + \epsilon_k \mathbf{I}\right)^{-1/2}\mathbf{1} \mathbf{I}.
\end{equation}
By means of employing this last inequality and the derivation of the previous appendix, a lower bound of the eigenvalues of $\widehat{\mathbf{Z}}_k$ can be established.
\section*{Acknowledgment}
The authors would like to thank Pantelis-Daniel M. Arapoglou for his fruitful comments and discussions.
%
%

\ifCLASSOPTIONcaptionsoff
  \newpage
\fi



\bibliographystyle{IEEEtran}
\bibliography{main}
\end{document}